\documentclass[11pt]{article}

\usepackage[margin=0.85in]{geometry}
\usepackage{setspace}
\emergencystretch=\maxdimen
\usepackage[T1]{fontenc}
\usepackage[utf8]{inputenc}
\usepackage{microtype}
\usepackage{textcomp}
\usepackage{amsmath}
\usepackage{amssymb}
\usepackage{mhchem}         

\usepackage{graphicx}
\usepackage{siunitx}
\usepackage{float}

\usepackage{authblk}
\usepackage{xcolor}
\usepackage[most]{tcolorbox}

\usepackage[
    backend=biber,
    style=nature,
    sorting=none,
    isbn=false,
    url=false,
    doi=true
]{biblatex}
\usepackage[hidelinks]{hyperref}

\title{\textbf{Atomic-Scale Characterization of Oxide Interfaces and
Superlattices Using Scanning Transmission Electron Microscopy}}

\author[1,2,3*]{Steven R. Spurgeon}
\author[1]{Renae N. Gannon}

\affil[1]{National Laboratory of the Rockies, Golden, CO 80401, USA}
\affil[2]{Metallurgical and Materials Engineering Department,
  Colorado School of Mines, Golden, CO 80401, USA}
\affil[3]{Renewable and Sustainable Energy Institute, University of Colorado Boulder,
  Boulder, CO 80309, USA}
\affil[*]{Corresponding author:
  \href{mailto:steven.spurgeon@nlr.gov}{steven.spurgeon@nlr.gov}}

\date{}

\begin{document}

\maketitle

\begin{abstract}
Scanning transmission electron microscopy (STEM) is a cornerstone of our understanding of oxide interfaces and
superlattices. No other technique provides the same level of insight into structure, chemistry, composition, and dynamics across as wide a variety of material systems. STEM imaging and diffraction, coupled with electron energy
loss (EELS) and energy-dispersive X-ray (EDS) spectroscopies, offer
unparalleled, high-resolution analysis of structure--property
relationships. In this chapter we highlight investigations into key
phenomena, including interfacial conductivity in oxide superlattices,
charge screening effects in magnetoelectric heterostructures, interface engineering in iron oxides, and the complex physics
governing atomic-scale chemical mapping. We also discuss emerging plasma preparation techniques and artificial intelligence-guided approaches to both ex situ and in situ microscopy. These studies illustrate how
unique insights from STEM characterization can be integrated with
other techniques and theory calculations to develop more predictive models for the behavior of functional oxides.

\vspace{0.5cm}
\end{abstract}

{\small\textit{Date: \today}}

\clearpage
\section{Introduction}

Scanning transmission electron microscopy (STEM) has long served as one of the most robust and versatile
platforms for materials characterization.\autocite{Hawkes2015,Krivanek2008,Muller2008,Bosman2007,Kimoto2007,Varela2004,Batson2002,Batson1993,Duscher1998,Browning1993,Muller1993}
Augmented with advanced aberration-correctors, modern electron
microscopes enable the simultaneous analysis of crystallography,
defects, chemistry, and composition at the atomic scale. Thin-film
oxide heterostructures, which possess tight coupling among spin,
lattice, and orbital degrees of freedom, represent ideal structures
for leveraging the power of STEM-based techniques. Precision
deposition methods, such as molecular beam epitaxy (MBE) and pulsed
laser deposition (PLD), permit the integration of systems with
disparate chemical, electronic, and magnetic properties, unlocking
emergent phenomena at
interfaces.\autocite{Hwang2012,Martin2010,Ramesh2007,Chambers2010b,Chambers2000}
Since these structures are typically synthesized in conditions far
from equilibrium, it is often difficult to predict their structure and
properties \textit{a priori}, necessitating detailed local investigation using STEM and related techniques. The information these probes provide can refine theory calculations that
can in turn be distilled into improved models for understanding the
behavior of these systems.

STEM techniques, encompassing imaging and diffraction, as well as
electron energy loss (STEM-EELS) and energy-dispersive X-ray
(STEM-EDS) spectroscopies, are uniquely suited to the analysis of
oxide heterostructures and interfaces. These materials generally
possess large lattice parameters that are compatible with slightly larger electron probes and the higher beam currents they permit, providing enhanced spectroscopic signals. The threshold for beam
damage in oxides is substantially higher than for many other
systems,\autocite{Egerton2004} permitting high-dose imaging and spectroscopy with minimal beam-induced damage. In addition, the transition metal cations in oxides generally exhibit sizable X-ray fluorescence yields that
facilitate rapid EDS composition mapping. Furthermore, investigations
using other X-ray core-level spectroscopies\autocite{de2008core} have
established a rich theoretical framework to interpret EELS fine
structure; experimental efforts are supported by \textit{ab initio}
calculations of electronic
structure\autocite{Ootsuki2014,Lopata2013,Ootsuki2011} as well as
multislice-based image and ionization map
simulations.\autocite{Allen2017,Allen2015,Kirkland2010,Cowley1957}

Our understanding of many oxide systems, with applications ranging from multiferroics to spintronics, has been greatly shaped by STEM characterization.\autocite{Gazquez2016,MacLaren2014,Varela2005} Using
STEM-EELS it is possible to visualize interfacial mixing, which is a
candidate mechanism for resolving the polar catastrophe at
polar/non-polar
interfaces.\autocite{Chambers2010,Muller2008,Nakagawa2006} Electronic
reconstructions at interfaces can also be directly
measured,\autocite{Cantoni2012,Shah2010a,Shah2010b} informing our
understanding of two-dimensional electron gas (2DEG) formation. Improvements in hardware and detectors combined with greater awareness of beam--specimen interactions in STEM-EDS\autocite{Allen2017,Allen2015,Kothleitner2014,Lugg2014,Forbes2012,DAlfonso2010,Oxley2007,Bosman2007,Findlay2005} and STEM-EELS\autocite{Allen2017,Gulec2015a,Xin2014,Forbes2012,Wang2008,Dwyer2008} have made atomic-scale analysis of interfaces commonplace.

Today, the challenge is no longer high-quality data acquisition but analysis and
throughput.\autocite{Spurgeon2021NatMater,Guinan2025APLML} Artificial intelligence
--- encompassing machine learning, automation, and robotics --- has begun to address
this challenge directly. Where we once manually extracted atomic-scale features from
individual images, machine learning approaches now operate across both data-rich and
data-poor regimes to quantitatively analyze interfaces at
scale.\autocite{Akers2021npjCM,Kalinin2022NatRevMeth,Ede2021MLST} This shift moves us
beyond isolated, anecdotal observations toward statistically representative models of
phase distributions, atomic-scale features, and interface
structures.\autocite{Nelson2021npjCM,Kalinin2022NatPhys,Ziatdinov2020npjCM,TerPetrosyan2025npjCM}
These models are particularly powerful for time-series data, where they can reveal
latent trends in phase transformations and guide the processing and synthesis optimization of functional
oxides.\autocite{Lewis2022npjCM} Such post-hoc approaches have now been integrated
into real-time microscope control, providing a ``brain'' for autonomous experimental
decision-making.\autocite{Olszta2022MicMic,Kalinin2021ACSNano,Kalinin2023npjCM} Instruments equipped
with computer vision, large language models, and reasoning agents can collect large
volumes of structured data to probe synthesis product distributions, map phase
transition trajectories, and identify dilute or rare
defects.\autocite{Guinan2025APLML,Guinan2026NatComm,Roccapriore2024SciAdv,Olszta2022MicMic}
Collectively, these capabilities allow us to map
structure--property--processing relationships, build statistical libraries of interface
configurations, and model materials breakdown under extreme
conditions.\autocite{Spurgeon2021NatMater,Kalinin2022NatPhys}

This chapter examines how advanced STEM techniques reveal the atomic-scale structure, chemistry, and dynamics of oxide interfaces and superlattices --- and how emerging AI-guided approaches are beginning to push that analysis toward statistical operation at scale. In Section~\ref{lco-sto} we discuss how STEM-EELS may be
used to probe cation intermixing and potential sources of interfacial
conductivity in model LaCrO$_3$ (LCO)/SrTiO$_3$ (STO)
superlattices. Section~\ref{lsmo} describes an approach to map
interfacial phases formed by screening of bound charge in
magnetoelectric La$_{1-x}$Sr$_{x}$MnO$_3$ (LSMO)/PbZr$_{x}$Ti$_{1-x}$O$_3$
(PZT) heterostructures; this study highlights how density functional
theory (DFT) calculations may be used to interpret EELS fine
structure measurements. Section~\ref{fe2o3} shows how information
from both STEM-EELS and STEM-EDS may be combined to improve the
design of high-quality Fe$_2$O$_3$/Cr$_2$O$_3$ superlattices for
photoelectrochemical water splitting. Section~\ref{lsco} addresses
commonly overlooked limitations of STEM-EDS mapping, illustrating the
dramatic effect of sample preparation on quantification of interfacial
mixing. Section~\ref{sec:pfib} describes advances in plasma focused
ion beam (PFIB) specimen preparation that have expanded the scope and
throughput of oxide interface characterization. Finally,
Section~\ref{sec:ai} reviews emerging AI-guided approaches, including few-shot machine learning for discovery, forecasting models for in-situ EELS, and multi-modal modeling of ion-induced degradation, that stand to transform experimental
workflows in the field.

\section{Cation Intermixing and Electronic Defects in LCO/STO
  Superlattices}
\label{lco-sto}

Perovskite oxides are among the most ubiquitous functional materials,
owing to their high degree of structural compatibility across
different stoichiometries and the wide range of emergent properties
at their surfaces and
interfaces.\autocite{Mannhart2010,Schlom2007} Extensive work has
focused on polar/non-polar interfaces, particularly in the archetypal
LaAlO$_3$ (LAO)/STO system,\autocite{Chen2024AdvElecMater, Ohtomo2004} but models
frequently assume a perfectly abrupt interface across which electron
transfer occurs. However, previous
studies\autocite{Qiao2011b,Chambers2010,Willmott2007a} have found
that intermixing is quite common in these materials, indicating that
extrinsic defects may play a more important role in mediating
interface conductivity than commonly thought. The success of this
past work suggests that asymmetrically terminated LCO/STO superlattices should similarly support a bulk-like built-in
potential gradient.\autocite{Comes2016,Chambers2011} While this
system is promising for separation of photoexcited electron--hole
pairs, the effect of intermixing and cation defects on the potential
remains poorly understood. The most rigorous characterization approaches
use Rutherford backscattering (RBS), X-ray absorption spectroscopy
(XAS), and X-ray photoelectron spectroscopy (XPS), which are either
volume-averaged or surface-limited; in contrast, STEM-EELS combines
fine structure information and high spatial resolution to locally
measure intermixing and electronic
defects.\autocite{Muller2008} A correlative approach, leveraging
multiple techniques, is essential to construct more accurate and
realistic interface models.

\begin{figure}[H]
\centering \includegraphics[width=0.85\textwidth]{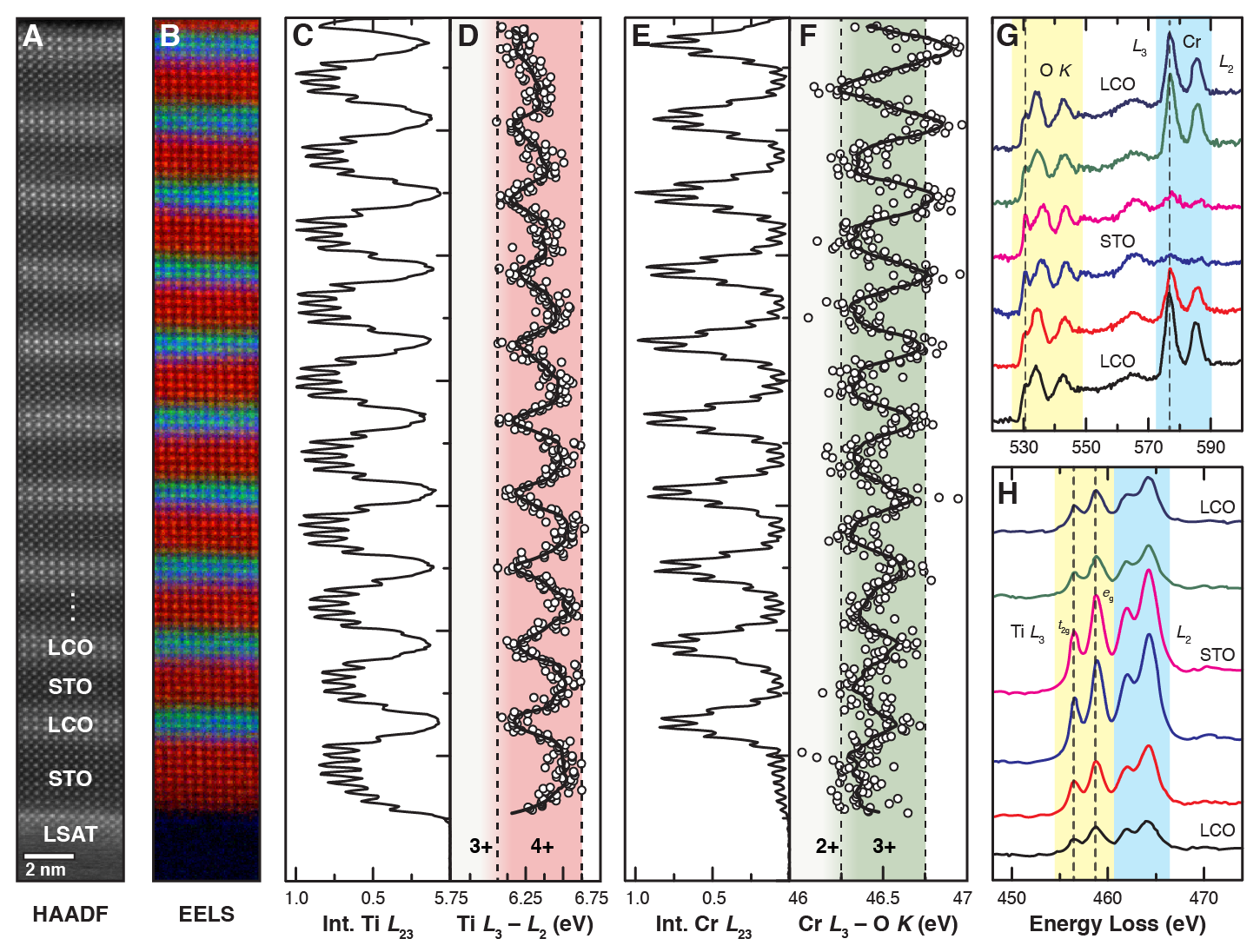}
\caption{\textbf{STEM-EELS analysis of cation intermixing and
defects in LCO / STO.} (\textbf{A}) Cross-sectional STEM-HAADF image of the
superlattice acquired during EELS mapping. (\textbf{B}) Composition
map of the Ti $L_{2,3}$, Cr $L_{2,3}$, and La $M_{4,5}$ (red, green,
and blue, respectively) edges constructed using multiple linear
least-squares fitting (MLLS). (\textbf{C}) Integrated Ti $L_{2,3}$
edge intensity profile. (\textbf{D}) Ti $L_3$--$L_2$ edge separation
profile used to estimate local Ti valence; values fall within the
Ti$^{4+}$ range (dashed lines). (\textbf{E}) Integrated Cr $L_{2,3}$
edge intensity profile. (\textbf{F}) O $K$--Cr $L_3$ edge energy loss separation used to estimate local Cr valence; values fall within the
Cr$^{3+}$ range (dashed lines). (\textbf{G}, \textbf{H}) O $K$/Cr
$L_{2,3}$ and Ti $L_{2,3}$ edge spectra, respectively, extracted
across the LCO/STO/LCO layers.
Reprinted with permission from reference~\cite{Comes2017}. Copyright
2017 American Chemical Society.\label{lco-sto_eels}}
\end{figure}

We have examined MBE-grown LCO/STO superlattices on
(LaAlO$_3$)$_{0.3}$(Sr$_2$AlTaO$_6$)$_{0.7}$ (LSAT) (001)
substrates.\autocite{Comes2017} While our previous STEM measurements
confirm a polarization induced by the alternation of
positively-charged [TiO$_2$]$^0$ $\vert$ [LaO]$^+$ and
negatively-charged [CrO$_2$]$^-$ $\vert$ [SrO]$^0$ interfaces, we
also find evidence for an unexpected conductivity in our
samples.\autocite{Comes2016} To probe the origin of this
conductivity, we have conducted extensive STEM-EELS mapping.
Figure~\ref{lco-sto_eels}A shows a cross-sectional STEM-HAADF image of the
superlattice, which reveals a high-quality
superlattice consisting of [6\,STO/3\,LCO]$_{\times 10}$ units. We
observe that the film is coherent and epitaxial, with no visible
misfit dislocations or secondary phases.
Figure~\ref{lco-sto_eels}B shows a composite STEM-EELS map of the
integrated La $M_{4,5}$, Cr $L_{2,3}$, and Ti $L_{2,3}$ signals,
highlighting the asymmetric termination at each block of the
superlattice. Figures~\ref{lco-sto_eels}C--D show line profiles
averaged in the film plane for the integrated Ti $L_{2,3}$ edge
intensity and Ti $L_3$ $t_{2g}$--$e_g$ crystal field splitting,
respectively; the latter may be used to investigate potential valence
changes, since the transition metal white-lines result from
excitations from spin--orbit split $2p_{3/2}$ and $2p_{1/2}$ states
to available $3d$ band states.\autocite{Varela2009} Similarly,
Figures~\ref{lco-sto_eels}E--F show line profiles averaged in the
film plane for the integrated Cr $L_{2,3}$ edge intensity and O $K$--Cr $L_3$ edge energy loss separation, respectively. These spectra
exhibit measurable Ti and Cr signals in all layers of the
superlattice, although the degree of Cr intermixing is less than that
of Ti. We have previously observed sizable Ti outdiffusion in LCO/STO
bilayers, which improves the thermodynamic stability of
the interface.\autocite{Colby2013}

With this intermixing established, we turn to potential valence changes and associated electronic defects. White-line
ratios, though commonly used to estimate local valence, are
challenging to compare to results from other groups, owing to
subjective differences in background subtraction as well as
position-sensitive effects.\autocite{Gulec2015a} Instead, we fit the
energy loss separation of the Ti $L_3$ and $L_2$ edges at each pixel
of the chemical map, generating the integrated line profile shown in
Figure~\ref{lco-sto_eels}D. We observe an average value across the
superlattice that falls within the literature range for a Ti$^{4+}$
state; however, the overall profile is tilted toward a slight Ti
valence reduction at the film surface, suggesting accumulation of excess La. The white-line ratio method is similarly
difficult to apply to the Cr $L_{2,3}$ edge because of its overlap
with the O $K$ continuum, so we instead consider the Cr $L_3$ to
O $K$ edge separation, which is a known indicator of local
valence.\autocite{Kaspar2016,ArevaloLopez2009} We again observe that
this quantity falls within the expected range of values for a
Cr$^{3+}$ state; however, we observe a slight reduction near the
interfaces with STO, indicating the possible compensation of
Ti$^{4+}$ cations in the nearby STO layers. A comparison to DFT
calculations for various defect scenarios shows that excess La can
give rise to additional carriers in the STO. Cr substitution for Ti
in the STO layer can lead to oxygen vacancy formation due to the
instability of octahedrally coordinated Cr$^{4+}$, suppressing
interfacial hole formation and preserving the built-in potential
across the superlattice.

Taken together, our results underscore the importance of correlative
high-resolution chemical and composition analysis alongside data from
volume-averaged techniques. STEM-EELS reveals a chemically realistic interface
that stands in stark contrast to the perfectly abrupt models typically
employed in theoretical calculations. The parameters
extracted from these measurements directly constrain
first-principles calculations, refining our understanding of the
complex interplay between structure and defects.

\section{Interfacial Phase Mapping of LSMO/PZT Magnetoelectric
  Heterostructures}
\label{lsmo}

Manganite-based heterostructures represent one of the most widely
studied oxide systems because of their intimate coupling among
structure, electronic, and magnetic order, which form the basis for
spintronic
devices.\autocite{Vaz2015,Wang2010,Wu2006,Dorr2006} Early reports of
colossal magnetoresistance
(CMR)\autocite{Dagotto2001,Coey1999} triggered extensive synthesis
and characterization efforts throughout the 1990s and early 2000s.
The magnetoelectric effect, which had previously been identified
only in a handful of monolithic compounds (\textit{e.g.},
Cr$_2$O$_3$\autocite{Astrov1960,Fiebig2005} and thin-film
BiFeO$_3$ (BFO)\autocite{Catalan2009}) was expanded to encompass a
new class of composite magnetoelectrics engineered to couple the
behavior of two or more phases at
interfaces.\autocite{Srinivasan2010} Composite structures based on
BaTiO$_3$ (BTO) and PZT soon exhibited magnetoelectric coefficients
orders of magnitude larger than in single-phase
materials.\autocite{Vaz2010c}

Coupling in these systems proceeds through several mechanisms, including charge screening, substrate-induced strain, and magnetic exchange.\autocite{Vaz2015} Much previous work has focused
on the presence of interface phases in these
structures,\autocite{Valencia2014,Huijben2008} which are thought to
mediate magnetoelectric
coupling,\autocite{Dong2013,Molegraaf2009} as well as interfacial
magnetization through dead layer
formation.\autocite{Jin2016,Vaz2015,Pradhan2013,Wang2010,Luo2008,Sun2008}
Particular attention has been paid to the LSMO/PZT
system,\autocite{Vaz2010c} which combines the manganite's high
ferromagnetic Curie temperature and spin
polarization\autocite{Cesaria2011,Moshnyaga2003} with the large
charge screening capabilities of the ferroelectric. Vaz \textit{et
al.} conducted the first \textit{in situ} poling measurements of ultrathin
LSMO on PZT during magneto-optical Kerr effect (MOKE) magnetometry,
demonstrating coupling mediated by a bound layer of interface
charge.\autocite{Vaz2011,Vaz2010a,Vaz2010b,Vaz2010c} Metallic
charge screening by Mn ions is thought to induce changes in the
effective local Mn valence, modifying both magnetic moment and spin
structure. While these early measurements provided important
information about coupling, they were unable to directly probe the
interface phases, chemistry, and spatial distribution of Mn
valence---all essential parameters needed to develop more complete
models of coupling.

Poling TEM samples \textit{in situ} presents significant practical challenges; instead, we developed an alternative substrate-induced self-poling technique.\autocite{Spurgeon2015,Yu2012} This method takes advantage
of the electrostatic boundary conditions of the substrate, which have
been shown to induce spontaneous unidirectional polarization of
PZT.\autocite{Spurgeon2014,Chen2013,Karthik2012,Yu2012,Afanasjev2001}
We first deposited a $\sim$12\,nm La$_{0.7}$Sr$_{0.3}$MnO$_3$ buffer layer on STO (001) by PLD, followed by a $\sim$37\,nm-thick PbZr$_{0.2}$Ti$_{0.8}$O$_3$ layer and a second $\sim$19\,nm-thick LSMO layer. This approach produces two different interface charge
states in a single sample, with the top layer in a hole charge
depletion state and the bottom layer in an accumulation state.

\begin{figure}[H]
\includegraphics[width=\textwidth]{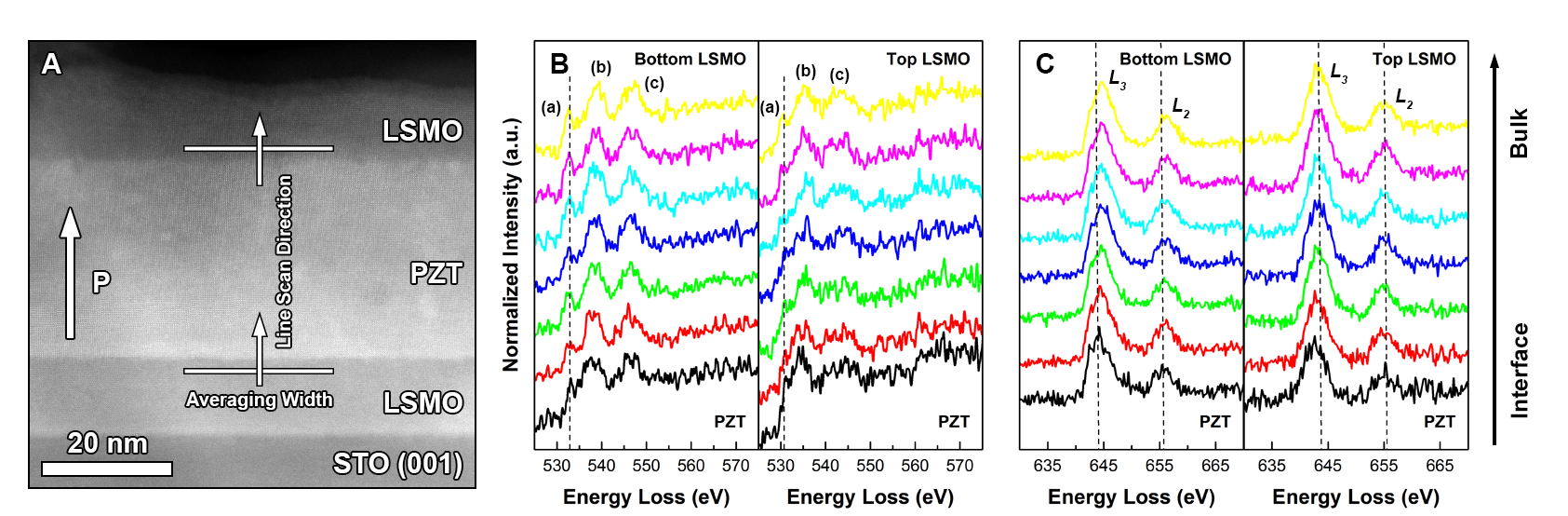}
\caption{\textbf{STEM-EELS mapping of the LSMO/PZT heterostructure.}
(\textbf{A}) Cross-sectional STEM-HAADF image of
the film structure, with the ferroelectric polarization direction
and linescan paths indicated. (\textbf{B}) O $K$-edge EEL spectra
collected plane-by-plane from the LSMO/PZT interface; \textbf{(a)},
\textbf{(b)}, and \textbf{(c)} label the pre-, main-, and
secondary-peak features, respectively. (\textbf{C}) Mn $L_{2,3}$-edge
EEL spectra across the same region.
Reprinted with permission from
reference~\cite{Spurgeon2015}. Copyright 2015. Rights managed by
Nature Publishing Group.\label{lsmo_profiles}}
\end{figure}

STEM-EELS allows us to measure spatially-resolved core loss spectra
to probe local bonding and Mn valence, which we can then relate to
the adjacent PZT
polarization.\autocite{Tan2012,Shah2010a,Varela2009}
Figure~\ref{lsmo_profiles}A shows a cross-sectional STEM-HAADF image
of the heterostructure, overlaid with a line scan along which O $K$
edge spectra were extracted in Figure~\ref{lsmo_profiles}B. The fine
structure of the edge arises from the dipole-allowed transition of
electrons from O $1s$ to unoccupied O $2p$-hybridized states; the three features
labeled (\textbf{a}--\textbf{c}) are associated with hybridized
states of O $2p$ with Mn $3d$, La $5d$/Sr $4d$, and Mn $4sp$
bands, respectively.\autocite{Varela2009} Clear shifts in the
spectral weight and position of the fine structure features are
evident across the LSMO/PZT interfaces, indicating local Mn valence
changes. Figure~\ref{lsmo_profiles}C shows the Mn $L_{2,3}$ edge,
which has been aligned to the commonly used position of the O $K$
edge near 532\,eV;\autocite{Varela2009} the guides to the eye mark a
distinct shift to lower $L_3$ energy onset in the vicinity of both
interfaces, denoting a Mn valence reduction. While this procedure is
not susceptible to energy drift errors, it is preferable to measure
the absolute energy of the Mn $L_{2,3}$ edge relative to a known
internal reference, such as the zero-loss peak or C $K$ edge if
possible, since the shape of the O $K$ edge can change greatly
depending on the compound.

\begin{figure}[H]
\centering \includegraphics[width=0.8\textwidth]{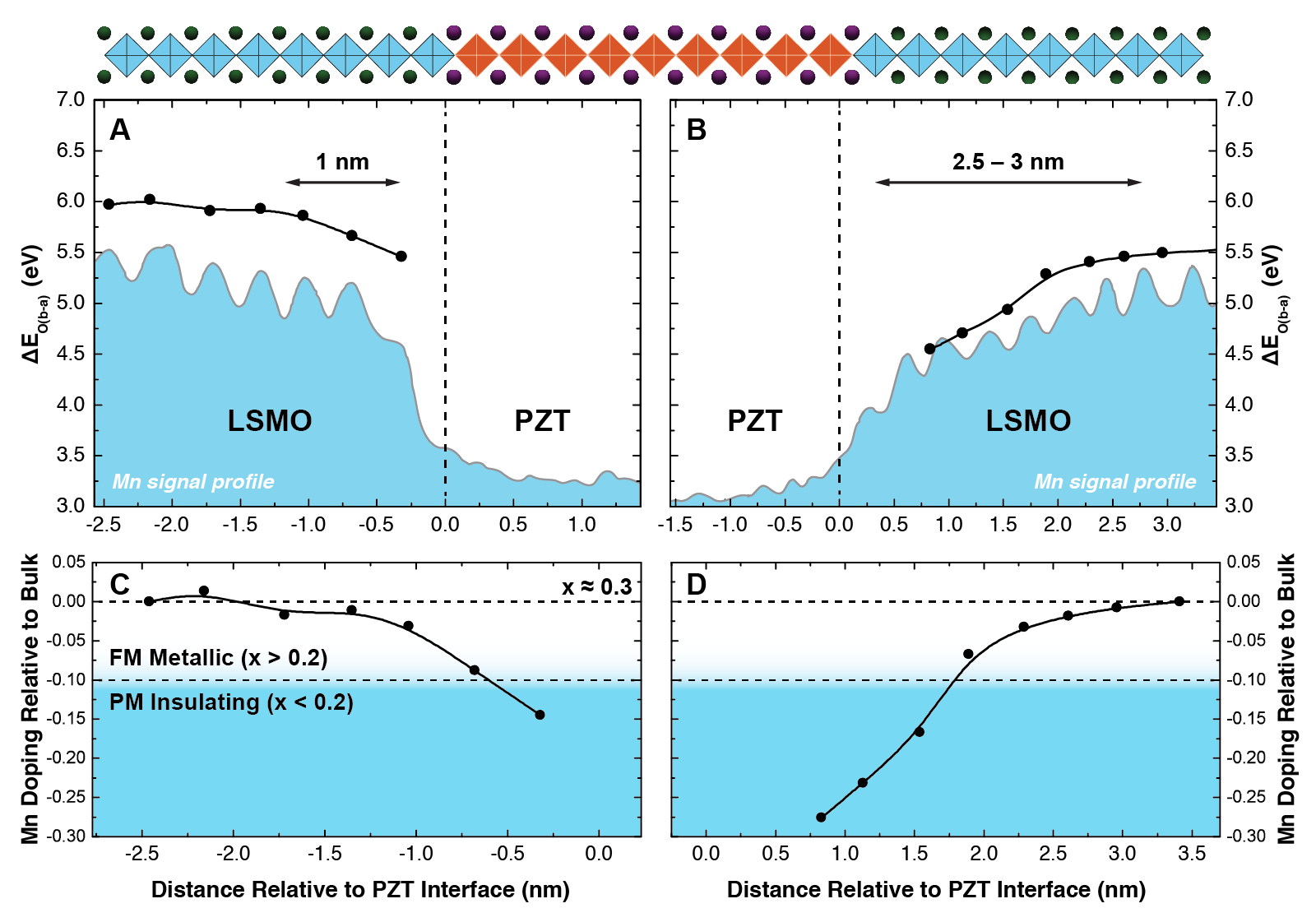}
\caption{\textbf{Spatially-resolved LSMO phase diagram.}
(\textbf{A}, \textbf{B}) O $K$ pre- to main-peak separation
($\Delta E_{\mathrm{O(b-a)}}$) in the vicinity of the PZT interface
for the bottom and top LSMO layers, respectively, overlaid with an
illustration of the heterostructure. (\textbf{C}, \textbf{D}) Map of
local Mn doping relative to bulk LSMO as a function of position
normal to the LSMO/PZT interface for the bottom (A) and top (B) LSMO
layers. The marked boundaries of the associated magnetic and
electronic phases are estimated from the bulk LSMO phase
diagram.\autocite{Zubko2011} Adapted with permission from
reference~\cite{Spurgeon2015}. Copyright 2015. Rights managed by
Nature Publishing Group.\label{lsmo_phasemap}}
\end{figure}

Figures~\ref{lsmo_phasemap}A--B show plots of the O $K$ pre- to
main-peak energy separation ($\Delta E_{\mathrm{O(b-a)}}$), which is
independent of energy calibration. Both interfaces exhibit a decrease
from a bulk value of 5.5--6\,eV to $<1.0$\,eV near the interface,
but the bottom LSMO shows this drop over 1\,nm versus 2.5--3\,nm for
the top interface. While these values agree qualitatively with the reference standards of Varela \textit{et al.},\autocite{Varela2009} direct, quantitative comparison to samples
measured on other microscopes is not advised because of differences
in peak fitting and microscope parameters. Instead, we compare our
values to DFT calculations to estimate the local variation in Mn
doping; this, in turn, may be related to the bulk LSMO phase diagram
to map the local phases present at both PZT interfaces.

Figures~\ref{lsmo_phasemap}C--D show the resulting spatially-resolved
LSMO phase diagram in the vicinity of the PZT. We find that the Mn
valence corresponds to the expected bulk La$_{0.7}$Sr$_{0.3}$MnO$_3$
(Mn$^{3.3+}$) doping up to three lattice planes away from the bottom
interface, where it drops by $\sim$0.15 over a nanometer. On the
other hand, the top interface exhibits a larger $\sim$0.26 drop over
a 2.5--3\,nm region, indicating an asymmetric coupling. Comparison to
the bulk LSMO phase diagram\autocite{Zubko2011} shows that the bottom
interface region is almost uniformly ferromagnetic and metallic to
the PZT interface, while the top region transitions to a broad
paramagnetic-insulating phase at the interface, which can explain the
observed asymmetry. This behavior agrees quite well with a model that
had been suggested for LSMO/BTO\autocite{Lu2012} but had never been
applied to LSMO/PZT. These first-of-their-kind measurements of local
phase gradients show that longer charge screening lengths can actually
be the byproduct of an electrostatic doping-induced phase transition.

\section{Interface Characterization of
  \texorpdfstring{Fe$_2$O$_3$/Cr$_2$O$_3$}{Fe2O3/Cr2O3}
  Heterostructures for Photocatalysis}
\label{fe2o3}

In Section~\ref{lco-sto} we discussed how STEM-EELS can reveal interfacial conductivity mechanisms and built-in potential profiles in photoactive superlattices. Hematite
($\alpha$-Fe$_2$O$_3$) is another system that has received
considerable attention because of its abundance, stability, and
suitable bandgap for
photocatalysis.\autocite{Shen2014} Nanostructured hematite-based
systems show promise for facilitating separation of electron--hole
pairs, while enhancing the effective surface area available for
photochemistry.\autocite{Kay2006} The superlattice layering of
Fe$_2$O$_3$ with Cr$_2$O$_3$ has been shown to exhibit favorable
band offset properties, leading to a sizable electrostatic potential,
which can further improve the separation of electron--hole
pairs.\autocite{Kaspar2013,Chambers2000a,Chambers1999} However, to
build and maintain this potential, we must strictly control layer
termination, composition, and intermixing, which may be verified using
STEM probes in conjunction with volume-averaged
techniques.

\begin{figure}[H]
\includegraphics[width=\textwidth]{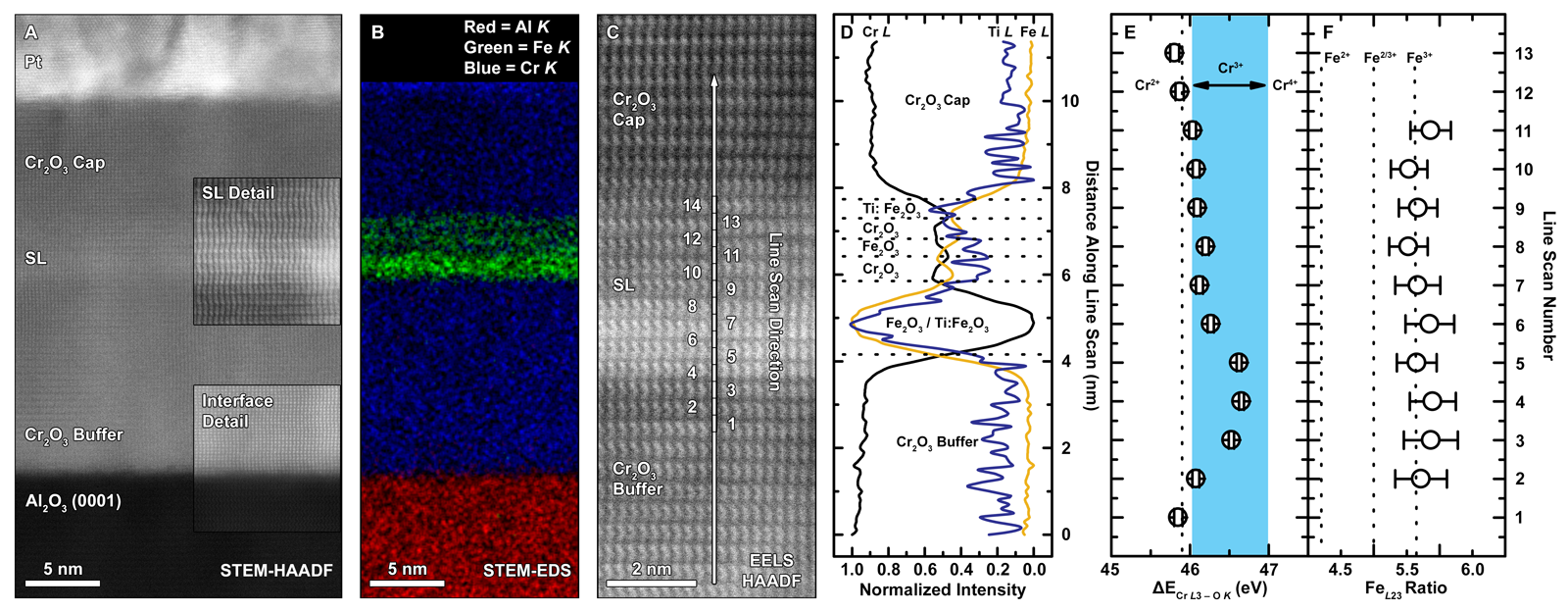}
\caption{\textbf{STEM analysis of the Fe$_2$O$_3$/Cr$_2$O$_3$
superlattice.} (\textbf{A}) Cross-sectional STEM-HAADF image of the
film, inset with high magnification images of the film--substrate
interface and superlattice. (\textbf{B}) STEM-EDS map of the Al $K$,
Fe $K$, and Cr $K$ peaks. (\textbf{C}) Radial-difference filtered
STEM-HAADF image of the region used for EELS measurements; a line
scan has been extracted and subsequent spectra have been integrated
in the marked windows. (\textbf{D}) STEM-EELS integrated signal
profiles extracted from the line scan region in C, processed using
principal component analysis, background subtracted, and normalized
to the Cr$_2$O$_3$ buffer layer. (\textbf{E}) O $K$--Cr $L_3$
edge energy loss separation used to estimate local Cr valence;
points fall within the Cr$^{3+}$ range (shaded region); error bars are
standard errors of Gaussian edge fits. (\textbf{F}) Fe $L_{2,3}$ white-line
ratios\autocite{VanAken2002a}; dashed lines indicate the literature range. Reprinted with permission from
reference~\cite{Kaspar2016}. Copyright 2016 John Wiley and
Sons.\label{fe2o3_mapping}}
\end{figure}

We have synthesized Fe$_2$O$_3$/Cr$_2$O$_3$ superlattices on
Al$_2$O$_3$ (0001) substrates using MBE; careful control of the
deposition ensures that each bilayer composition provides the largest
conduction and valence band offsets, resulting in a large built-in
electric field across the superlattice.\autocite{Kaspar2016}
Figure~\ref{fe2o3_mapping}A shows a STEM-HAADF image of the overall
film structure, inset with high magnification views of the
film--substrate interface and superlattice. We observe that the
interface with the Al$_2$O$_3$ substrate is coherent, uniform, and
defect-free. Furthermore, all layers of the superlattice are
preserved, albeit with some undulation due to the large lattice
mismatch of $\Delta a/a = 4.2\%$.

While the STEM-HAADF image shows a clear contrast between the film
and substrate, the similar atomic number ($Z$) of the film layers
necessitates the use of spectroscopy to directly resolve the
superlattice components. Although STEM-EDS offers poorer energy
resolution ($\sim$10\,eV) than STEM-EELS ($<$0.1\,eV), it measures a wide range of ionization energies
simultaneously, making it well-suited to the analysis of
metal-oxides.\autocite{Williams2009}
Figure~\ref{fe2o3_mapping}B shows a composite STEM-EDS map of the
Al $K$, Fe $K$, and Cr $K$ edges; this map confirms the overall
composition and geometry of the film layers. To resolve the electronic fine structure, we
acquired a STEM-EELS line scan from the region in
Figure~\ref{fe2o3_mapping}C; the line scan geometry limits cumulative electron dose, thereby preserving the integrity of the fine structure.
Figure~\ref{fe2o3_mapping}D shows the power-law background-subtracted
integrated Fe $L_{2,3}$, Cr $L_{2,3}$, and Ti $L_{2,3}$ edge
signals, normalized to their concentration in the Cr$_2$O$_3$ buffer
layer. The profile confirms the superlattice configuration in
agreement with the STEM-EDS results, but the added spatial resolution
shows that there is intermixing over $\sim$0.5\,nm near the bottom
and top of the superlattice.

Turning to the electronic fine structure, we next explore systematic
changes in the energy separation of the O $K$ and Cr $L_3$ edges, as
was done in Section~\ref{lco-sto}; using this approach, we avoid
deconvolving the tail of the O edge from the onset of the Cr edge.
Figure~\ref{fe2o3_mapping}E shows the measured energy separation,
overlaid with the range of values corresponding to a Cr$^{3+}$-like
state extracted from the literature.\autocite{ArevaloLopez2009} We
observe a slight increase toward more Cr$^{4+}$-like character near
the bottom of the superlattice, but the overall valence falls
squarely within the bounds of the Cr$^{3+}$-like state; this
behavior may reflect intermixing with the Fe layer, as is evident in
Figure~\ref{fe2o3_mapping}C. To gain insight into the valence of the
Fe, we have conducted measurements of the Fe $L_{2,3}$ white-line
ratio according to the method of van Aken \textit{et
al}.,\autocite{VanAken2002a} as shown in
Figure~\ref{fe2o3_mapping}F. Although we again emphasize that
quantitative comparison to other results in the literature is
difficult, owing to differences in background subtraction and peak
fitting methodologies, we find that the ratio falls within the range
of values expected for an Fe$^{3+}$-like
state.\autocite{Cave2006c,VanAken2002a}

This study confirms that it is possible to synthesize high-quality
Fe$_2$O$_3$/Cr$_2$O$_3$ superlattices using MBE. While STEM-EELS
indicates some intermixing near the top and bottom of the
superlattice, the overall chemistry of the Fe and Cr layers is
preserved. In concert with XPS and related
characterization,\autocite{Kaspar2016} we find that the favorable
band alignment and enhanced photoconductivity of these materials may
facilitate improved photoelectrochemical water splitting.

\section{Challenges to Quantification of Atomic-Scale STEM-EDS Maps}
\label{lsco}

Chemical mapping using STEM-EELS and STEM-EDS offers several advantages compared to
non-spectroscopic imaging. In particular, these techniques are
directly sensitive to the elemental composition of a sample, whereas STEM-HAADF contrast, while $Z$-sensitive, is less readily converted to quantitative composition. Furthermore, the enhanced
energy resolution in EELS allows the operator to directly interrogate
electronic fine structure and properties such as valence and bonding.
However, strong interactions between the incident electron beam and
lattice can lead to serious artifacts that preclude straightforward
analysis of atomic-scale chemical maps. Effects such as beam
broadening and channeling\autocite{Oxley2007} are particularly
troublesome along low-order zone axes, resulting in coupling between
the probe and atomic columns.\autocite{Lugg2014} Thermally scattered
electrons can influence the ionization signal in both STEM-EDS and
STEM-EELS,\autocite{Forbes2012} as well as introduce image contrast
reversals in STEM-EELS due to off-column probe
channeling.\autocite{Wang2008,Oxley2007} Although first-principles
calculations have been able to model the Bremsstrahlung background in
STEM-EDS,\autocite{Williams2009} resulting in a simpler signal
integration process,\autocite{DAlfonso2010} channeling and probe
broadening through the specimen still make it difficult to accurately
quantify unknown interfaces.\autocite{Kotula2012}

\begin{figure}[H]
\centering{\includegraphics[width=0.6\textwidth]{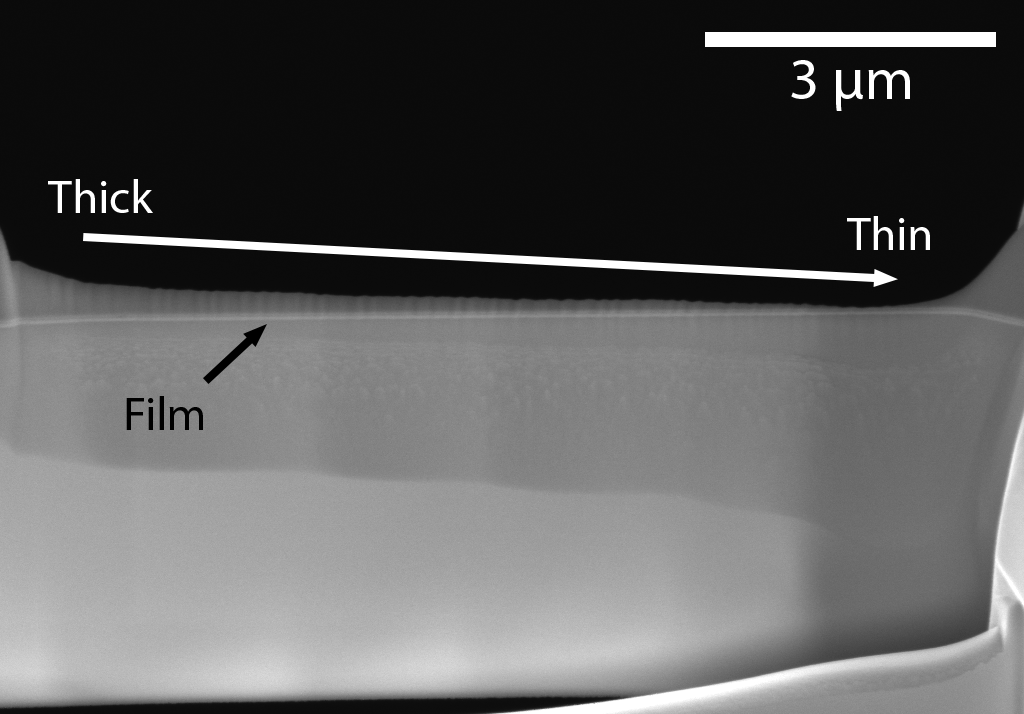}}
\caption{\textbf{STEM sample preparation.} Cross-sectional secondary
electron image of the polished STEM lift-out, with the gradient from
thick to thin foil marked by the arrow. The bright horizontal band
corresponds to the location of the film--substrate
interface. Reprinted with
permission from reference~\cite{Spurgeon2017Chapter}. Copyright 2018 Elsevier.\label{lsco_sem}}
\end{figure}

To better understand sources of error in atomic-scale STEM-EDS
mapping of interfaces, we have characterized a model heterostructure
that shows potential use for a transparent, perovskite-based $p$--$n$
junction.\autocite{Spurgeon2017} This system consists of a
30\,nm-thick La$_{0.88}$Sr$_{0.12}$CrO$_3$ ($p$-type LSCO)
thin-film deposited onto a 0.1\,wt\% Nb:STO (001) ($n$-type Nb:STO)
substrate using MBE.\autocite{Zhang2015} The wedge-shaped sample
shown in Figure~\ref{lsco_sem} was prepared using a modified focused
ion beam (FIB) lift-out procedure to produce a range of
cross-section thicknesses. STEM-EDS maps were subsequently acquired
from different regions spanning 28, 33, 50, 66, 70, and 75\,nm
thick, according to zero-loss peak measurements.

\begin{figure}[H]
\includegraphics[width=\textwidth]{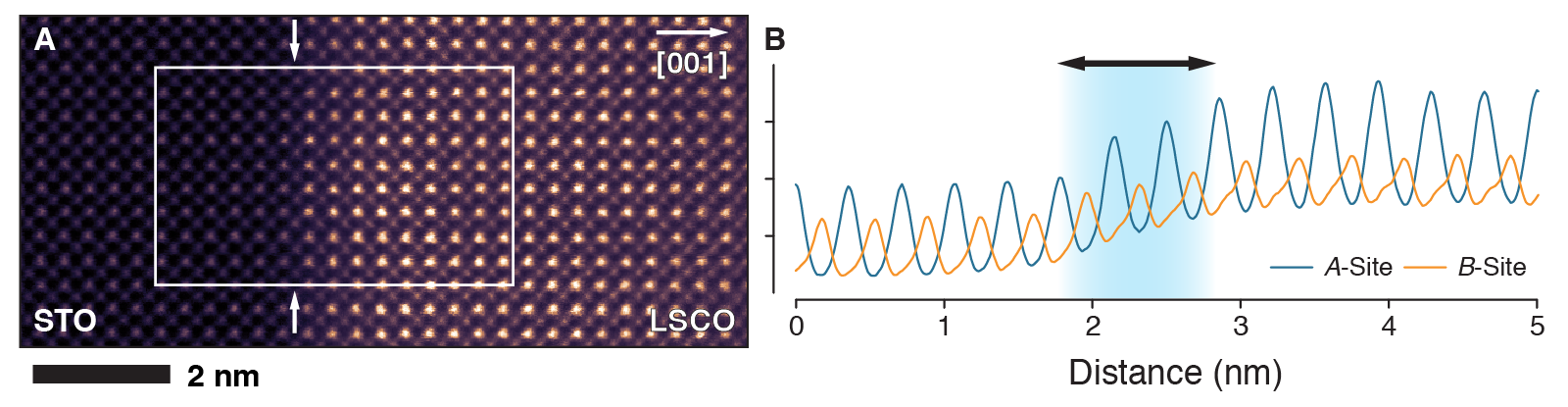}
\caption{\textbf{LSCO/STO interface characterization.}
(\textbf{A}) Representative cross-sectional STEM-HAADF image of the
LSCO/STO interface taken along the [100] zone-axis. (\textbf{B})
Average of 10 $A$- and $B$-site profiles taken across the interface,
with the shaded region indicating the interface
position. Reprinted with
permission from reference~\cite{Spurgeon2017Chapter}. Copyright 2018 Elsevier.\label{lsco_haadf}}
\end{figure}

Figure~\ref{lsco_haadf}A shows a representative cross-sectional
STEM-HAADF image of the heterostructure. The film is epitaxially registered to the substrate, with no obvious
defects or misfit dislocations.\autocite{Colby2013} The contrast
clearly changes across the film--substrate interface, resulting from
the differing atomic numbers of the cation species, as well as a
varying background contribution that is known to depend on sample
thickness\autocite{Klenov2006} and oxygen content. Even in the case
of atomically abrupt interfaces, a finite sample thickness gives rise
to an intermediate background different from either component of the
junction.\autocite{Klenov2006} While zero-loss measurements of the
sample show minimal thickness fluctuations across the interface, we
still observe a sloping background variation, as shown in the
extracted line profile in Figure~\ref{lsco_haadf}B. Such a
background hinders accurate quantification of intermixing, since it
is possible to define the interface width in several ways: (1)
relative to the changing background, (2) relative to the signal
after background subtraction, or (3) relative to the column ratios
within each layer.\autocite{Robb2012}

\begin{figure}[H]
\includegraphics[width=\textwidth]{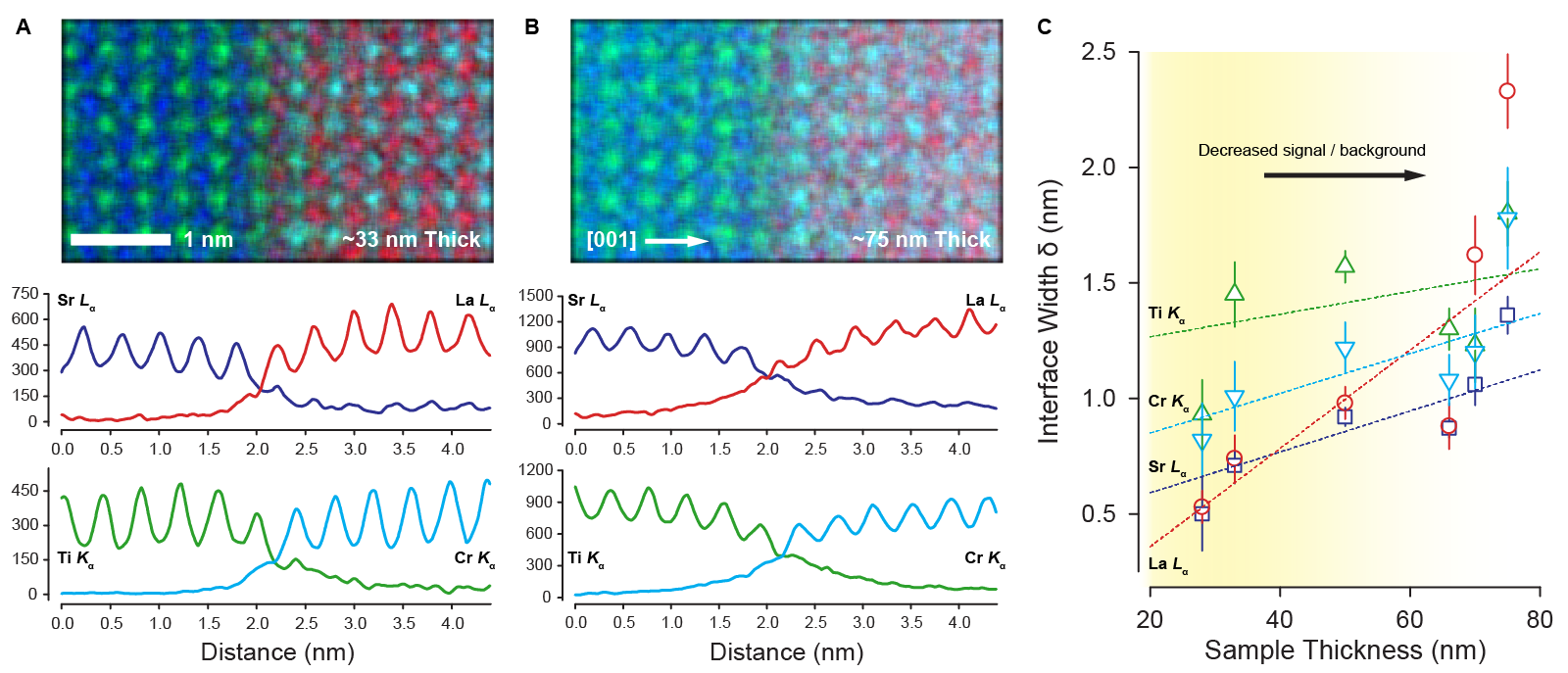}
\caption{\textbf{Thickness dependence of STEM-EDS mapping.}
(\textbf{A}--\textbf{B}) Composite STEM-EDS maps and
corresponding $A$- and $B$-site net X-ray count line profiles for
$\sim$33 and $\sim$75\,nm-thick STO/LSCO interfaces, respectively;
line profiles have been averaged in the plane of the maps.
(\textbf{C}) Interface width as a function of sample thickness for
each species. Adapted with permission from
reference~\cite{Spurgeon2017}. Copyright 2017 Microscopy Society of
America.\label{lsco_eds}}
\end{figure}

Since the cation species are chemically similar, STEM-EDS provides a more directly interpretable measure of interdiffusion; Figures~\ref{lsco_eds}A--B show composite STEM-EDS maps and corresponding averaged line profiles for Sr $L_\alpha$ (1.806\,keV),
La $L_\alpha$ (4.650\,keV), Ti $K_\alpha$ (4.508\,keV), and
Cr $K_\alpha$ (5.412\,keV) taken from $\sim$33 and $\sim$75\,nm-thick
regions of the sample, respectively. We define the apparent interface
width ($\delta$) as the difference of 90\% and 10\% of the signal
maxima from a logistic fit to either side of the interface. These
profiles clearly reveal some intermixing, particularly of Ti and Cr,
as we have observed in other samples.\autocite{Comes2017} Moreover,
there are noticeable differences in the interface width between the
two regions, as well as a higher background in the thicker part of
the foil. We repeated this procedure for a range of foil
thicknesses, as shown in Figure~\ref{lsco_eds}C. We observe a clear
trend toward more diffuse interfaces in thicker samples, ranging from
2--5 unit cells of artificial broadening depending on the species.
We find that $\delta_\textrm{La}$ increases to $\sim$1.8\,nm moving
from 28 to 75\,nm-thick regions; similarly, $\delta_\textrm{Sr}$ and
$\delta_\textrm{Cr}$ increase by $\sim$0.86 and $\sim$0.96\,nm,
respectively, while $\delta_\textrm{Ti}$ increases by $\sim$0.87\,nm.
To confirm that these changes are not due to the presence of
substrate steps, we have conducted atomic force microscope (AFM)
measurements of the STO surface prior to growth; our measurements
show that at most one step is present along the beam direction
for the measured foil thicknesses, ruling out steps as a source of
the artificial interface broadening. These results indicate that measured interface widths depend strongly on both sample preparation and the choice of ionization edge, with each exhibiting distinct thickness-dependent trends.

We have conducted multislice ionization
simulations\autocite{Spurgeon2017} that confirm a consistent trend
toward artificially broadened interfaces in thicker regions. We find
that a foil thickness of only 40--50\,nm is sufficient to introduce
multiple unit cell measurement errors arising from complex beam
channeling effects, as well as the delocalization character of each
edge. We strongly recommend that rigorous sample preparation routines
be employed to achieve the thinnest foils, with the best results
obtained below 25--30\,nm. However, at these thicknesses it is
difficult to achieve a reasonable ionization signal, necessitating
the use of lower accelerating voltages (increasing the ionization
cross-section), large-area EDS detectors, or advanced drift
correction routines to enhance signal-to-noise.\autocite{Wenner2017,Jones2015}
Regardless, our results indicate that atomically resolved chemical maps require accompanying simulations to fully account for complex beam--specimen interactions.

\section{Plasma Focused Ion Beam Specimen Preparation for Interface Characterization}
\label{sec:pfib}

As already discussed, meticulous sample preparation is key to achieving accurate, quantitative analysis of interfaces. Conventional Ga liquid metal ion sources (LMIS) have long been the standard FIB source for site-specific TEM specimen preparation. However, there are significant drawbacks to preparing oxide materials with Ga-FIB, as it is widely known to accumulate at grain boundaries and induce phase changes and defects in materials. High-energy Ga$^+$ ions induce deep collision cascades, leading to amorphization, vacancy generation, and implantation --- ultimately producing unwanted chemical doping and defects.\autocite{roadmap_for_FIB} These beam-induced artifacts are particularly destructive for oxide heterostructures, where emergent properties are highly sensitive to structural defects and interface chemistry. For systems like NiO/Ga$_2$O$_3$, implanted Ga may be indistinguishable from the substrate, inhibiting analysis of the native interface chemistry and structure.

Recent developments and commercialization of plasma focused ion beam (PFIB) and gas field ion source (GFIS) systems have demonstrated the ability to overcome many limitations and artifacts of Ga-FIB preparation or polishing of site-specific lamellae. Both technologies are capable of using noble gases (commonly Xe, Ar, Ne, He) as the ion species, enabling inert preparation and processing, where PFIB systems --- which can also operate with reactive species such as O and N --- are more commonly used for sample preparation due to higher sputter yields and throughput. The mass difference among ion species governs interaction volume: heavier Xe$^+$ ions have shallower penetration depth and produce thinner amorphous sidewall damage compared to Ga$^+$ ions.\autocite{zhong2021} Conversely, lighter species such as Ar$^+$ have larger penetration depth than Ga$^+$ ions. Experimentally, implantation of ions can extend much deeper than simulations predict due to ion channeling effects in crystalline materials.\autocite{multi_ion_PFIB}

\begin{figure}[H]
    \centering
    \includegraphics[width=0.8\textwidth]{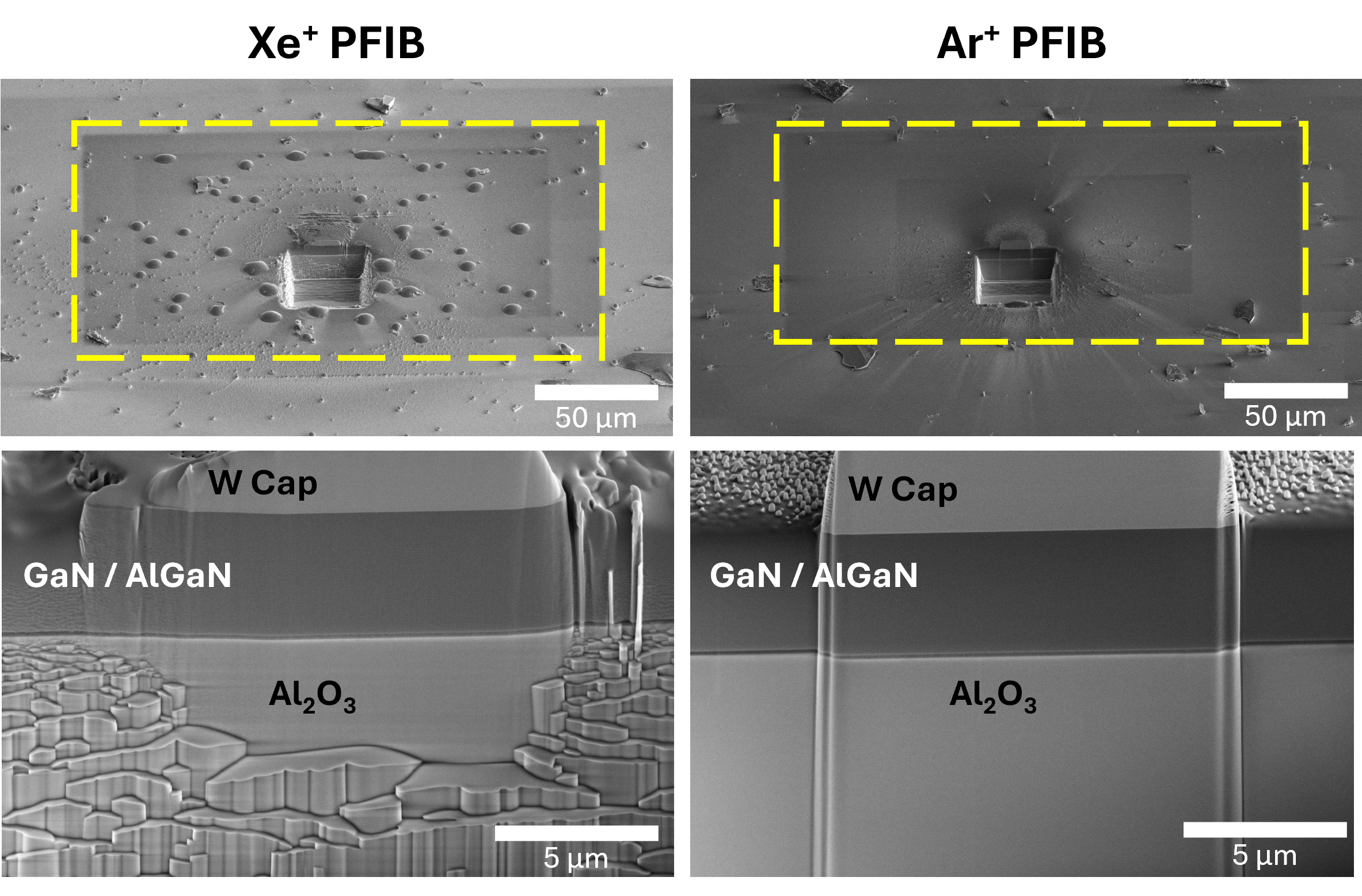}
    \caption{\textbf{Surface textures induced by different ion species.} Xe$^+$- and Ar$^+$-PFIB-exposed regions of AlGaN/GaN on Al$_2$O$_3$ are highlighted in yellow (top row). Cross-sections of each prepared specimen are shown below (bottom row). Only the Xe$^+$-exposed surface exhibited bubbling; Xe$^+$ also produced terrace-like curtaining in Al$_2$O$_3$ not observed with Ar$^+$.}
    \label{fig:PFIB comparison}
\end{figure}

While chemical artifacts such as doping or accumulation can be minimized or eliminated, careful attention must be given to beam conditions and ion-sample interactions to minimize amorphization and defect generation. For example, using accelerating voltages $< 16$~kV for species such as Ar$^+$ for thinning results in interaction volumes that are comparable to 30~kV Ga$^+$ and Xe$^+$, and minimizes damage to the specimen. Ion species selection is also a critical consideration since the ion-sample interactions of these relatively new ion species for sample preparation are not yet understood or widely studied. An example of cross-sections in AlGaN/GaN on Al$_2$O$_3$ prepared by 30~kV Xe$^+$ and Ar$^+$ PFIB are shown in Figure~\ref{fig:PFIB comparison}, where only the surface exposed to Xe$^+$ PFIB resulted in bubbling of the surface. Additionally, the Al$_2$O$_3$ formed terrace-like curtaining artifacts when sectioned with Xe$^+$ which were not present when sectioning with Ar$^+$. Materials and samples that exhibit this behavior should be prepared with Ar$^+$ PFIB $< 30$~kV to minimize damage and optimize surface quality.

Despite these advantages, achieving ultra-thin electron-transparent lamellae ($< 50$~nm) is challenging with PFIB. PFIB instruments generally exhibit broader beam tails, which demand more careful management during specimen thinning, particularly at lower accelerating voltages where precise end-pointing becomes critical.\autocite{suzy_vitale} There are several strategies for preparing high quality, ultra-thin lamellae through PFIB. Typically, best practices for PFIB lamella preparation involve depositing thicker protective caps than those used traditionally with Ga-FIB, since broader beam tails will erode the protective cap during thinning. The choice of thinned window area is equally important: a targeted, narrow window consistently outperforms a large-area approach for achieving ultra-thin lamellae. In oxide interfaces for example, this means strategically using over- and under-tilt angles that target thinning the interface, rather than the full height of the specimen. For traditional top-down thinning, this requires a very thick protective cap and shallow tilt angles.

\begin{figure}[H]
\includegraphics[width=\textwidth]{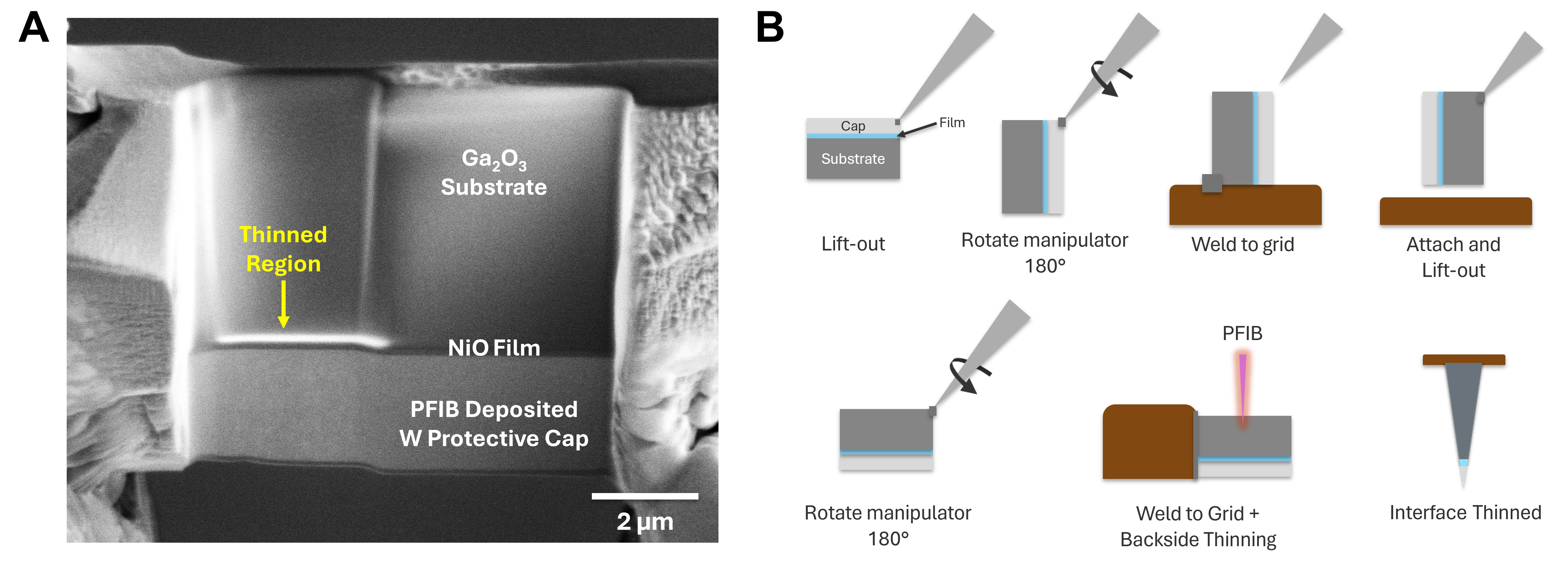}
\caption{\textbf{Inverted lift-out preparation workflow.} (\textbf{A}) A thinned window ($<50$\,nm) on an inverted NiO/Ga$_2$O$_3$ TEM specimen prepared by Xe and Ar PFIB. (\textbf{B}) Inverted specimen workflow for a rotation-capable nanomanipulator.}
\label{inverted_liftout}
\end{figure}

Another approach to targeted thinning is using backside milling, where the sample is thinned from the substrate side. While backside thinning is often more time and labor intensive, requiring rotations and re-attachment of the specimen to the TEM grid, there are several benefits. First, protective cap erosion is less of a concern and protective cap depositions do not need to be as thick as they would be for top-down thinning. Next, high over- and under-tilt angles can be used to create a thin wedge at the bottom of the sample to obtain an ultra-thin region where a film/substrate interface would be, for example. Redeposition and curtaining are less of a consideration when using this approach, compared to targeting from a top-down geometry. This also provides mechanical stability, as the top of the sample (substrate) can remain thick. In any thinning strategy, thinning narrower width windows ($< 3~\mu$m) helps with mechanical stability and minimizes warping of ultra-thin specimens. An inverted NiO/Ga$_2$O$_3$ specimen and workflow for inverted sample preparation are shown in Figure~\ref{inverted_liftout}. While traditional sample preparation procedures need to be modified for PFIB or noble gas ion species, higher throughput and quality are achievable. Multi-species approaches, including bulk preparation and thinning with Xe$^+$ followed by final low-voltage ($< 3$~kV) Ar$^+$ polishing\autocite{vasquez2025}, provide the highest throughput while minimizing damage and beam artifacts.

\section{Toward AI-Guided Characterization of Oxide Interfaces}
\label{sec:ai}

\begin{figure}[H]
\centering \includegraphics[width=0.85\textwidth]{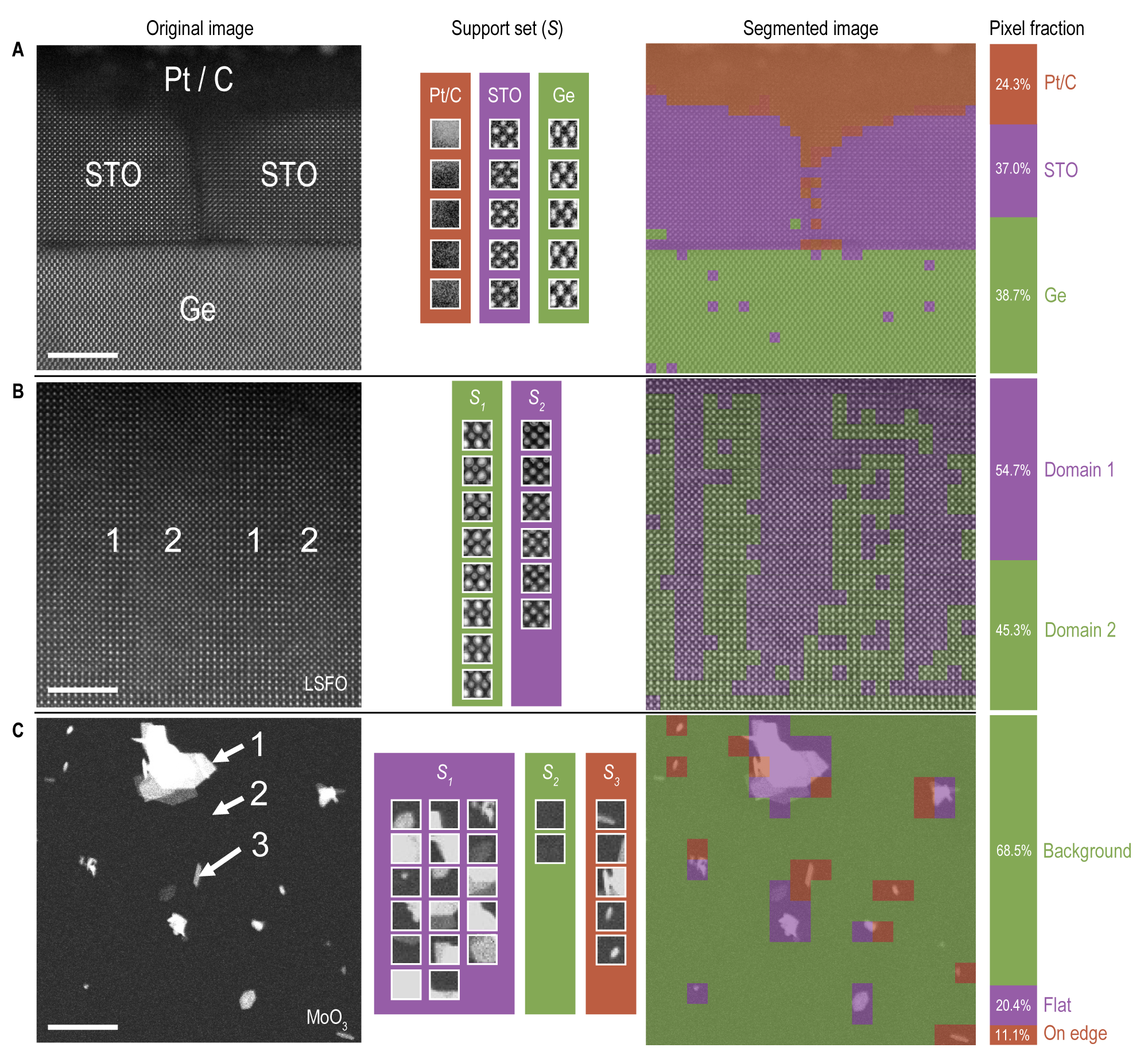}
\caption{\textbf{Few-shot machine learning enables rapid, label-efficient segmentation of oxide microstructures.} (\textbf{A})~Segmentation of an epitaxial SrTiO$_3$/Ge heterostructure interface, (\textbf{B})~phase separation in La$_{0.8}$Sr$_{0.2}$FeO$_3$ thin films, and (\textbf{C})~MoO$_3$ nanoparticle morphologies, shown with original images (left), support sets (center), and segmented outputs with quantified phase fractions (right). The approach generalizes across material systems and length scales without retraining, enabling high-throughput microstructural analysis for emerging autonomous characterization platforms. Adapted with permission from reference~\cite{Akers2021npjCM} under CC-BY 4.0 license. Copyright 2021 the Authors. \label{few_shot}}
\end{figure}

As aberration-correction and large, high-speed detectors have become mainstream, the microscopy community has increasingly begun to struggle with decision-making on large volumes of multimodal data. Purely human-in-the-loop workflows, once sufficient to analyze a handful of interface images or chemical maps, now struggle to keep pace with this data deluge. We face a major bottleneck in labeling and interpreting this data, leading us to expend considerable time and effort analyzing just a fraction of the data at our disposal. Conventional supervised ML, which could assist with this process, typically requires substantial amounts of model training data --- something largely unattainable for most oxide interfaces. To address this, we have extensively implemented sparse data approaches based on few-shot machine learning, in which models learn by comparison .\autocite{Akers2021npjCM,Doty2022CompMatSci} As shown in Figure~\ref{few_shot}, these models can perform classification of atomic motifs, phases, and defects from as few as 5--8 labeled sub-images (termed a ``support set'') without retraining across material systems. This approach is well-suited to rapid detection of features in microscopy data, where we often have limited intuition and need to triage data in discovery scenarios.

Figure~\ref{few_shot}A--C demonstrates the approach across three representative material systems: an epitaxial SrTiO$_3$/Ge heterostructure interface (panel A), phase separation in La$_{0.8}$Sr$_{0.2}$FeO$_3$ thin films (panel B), and MoO$_3$ nanoparticle morphologies (panel C). In each case, the same model generalizes across length scales and materials without modification; only the support set composition changes. Using this approach, it is possible to rapidly quantify the distribution of phases and defects across interfaces---a key step in moving from qualitative imaging to statistical understanding of synthesis products. Moreover, it is possible to understand how atomic motifs resulting from mosaicity and oxygen deficiency arise as a function of synthesis condition. This approach enables a broader autonomous characterization paradigm, in which label-efficient segmentation is a key enabler for real-time screening of materials.

In other studies, probing the dynamic pathways for defect formation and phase transformation in situ is essential, rather than characterizing synthesis products post hoc. Modern instruments are capable of exquisite control of electron beam and sample environment, allowing us to collect dynamic data at the nanoscale. However, traditional ML approaches struggle to provide actionable, real-time feedback during such an experiment and we often miss key features of a phase transformation due to slow detection. To address this, we have developed a long short-term memory forecasting model for materials dynamics called EELSTM.\autocite{Lewis2022npjCM} This is a qualitatively different capability than post hoc analysis, in that it can predict the future trajectory of a spectral signal in time-series EELS data. Such data are critical to understanding beam-sensitive oxides, mitigating unwanted artifacts, and probing the nature of electronic phase transitions induced by oxygen vacancy formation.

\begin{figure}[H]
\centering \includegraphics[width=0.7\textwidth]{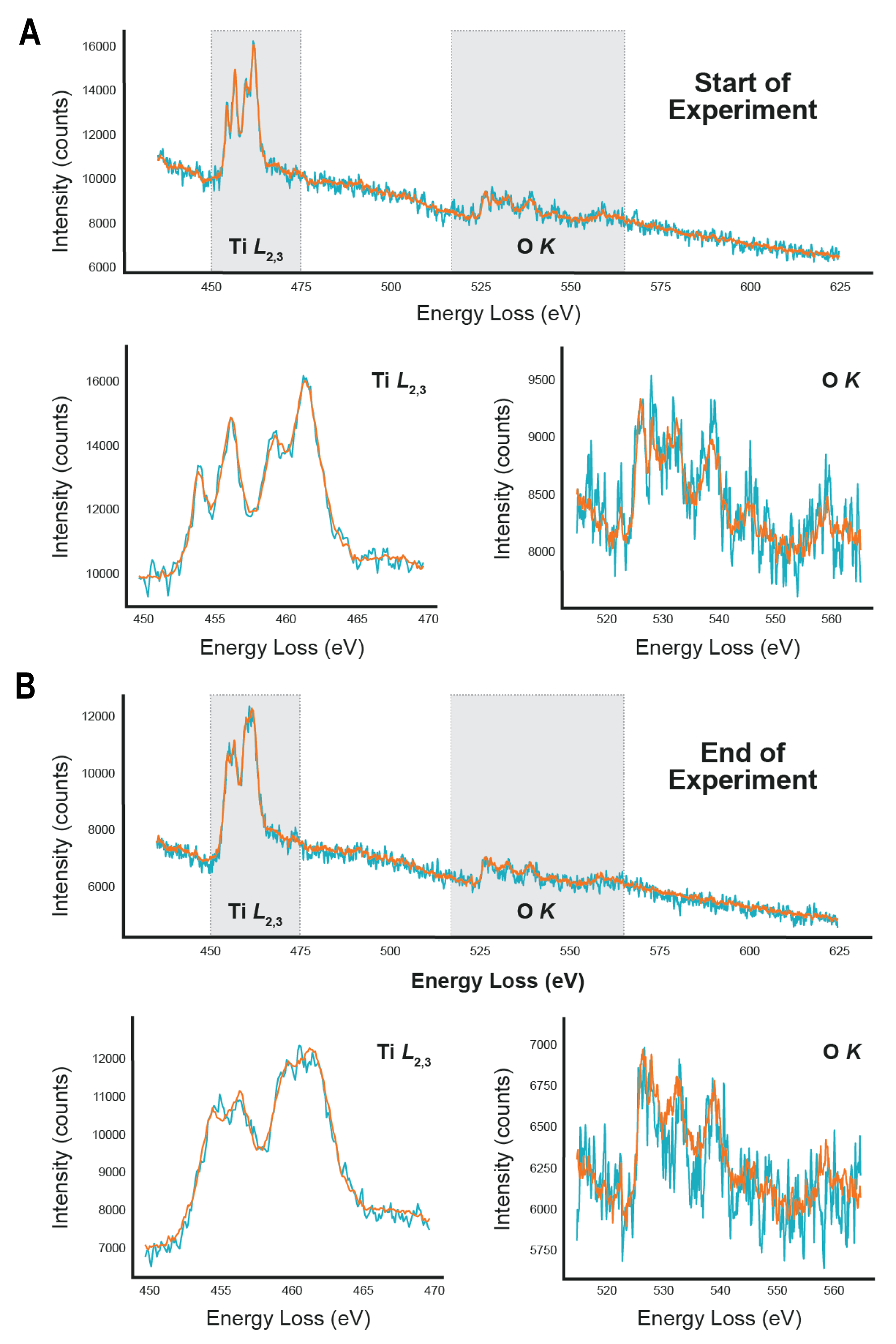}
\caption{\textbf{Recurrent neural network forecasting captures electron beam-driven valence changes in a perovskite oxide.} A long short-term memory (LSTM) model trained on in situ EELS time-series data predicts the trajectory of a crystalline-to-amorphous phase transition in SrTiO$_3$ driven by electron beam reduction. (\textbf{A})~At $t \approx 15$\,s, the predicted spectrum (orange) reproduces the Ti$^{4+}$-dominated fine structure, including the $t_{2g}$/$e_g$ crystal-field splitting and O~$K$ edge. (\textbf{B})~At $t \approx 60$\,s, the model tracks the emergence of Ti$^{3+}$ character via merging of the $L_3$/$L_2$ white lines and progressive O~$K$ edge flattening. Ground-truth spectra are shown in blue. Adapted with permission from reference~\cite{Lewis2022npjCM} under CC-BY 4.0 license. Copyright 2022 Battelle Memorial Institute. \label{LSTM}}
\end{figure}

Figure~\ref{LSTM} demonstrates how the EELSTM model works for in situ electron beam-induced reduction of SrTiO$_3$. The model is first trained on evolving Ti $L_{2,3}$ and O $K$ edge fine structure as the specimen transitions from crystalline to amorphous under electron beam reduction. These edges are known leading indicators of electronic state change and are highly sensitive to the presence of direct Ti reduction and oxygen vacancies. Figure~\ref{LSTM} shows the state of the two edges at the beginning of an experiment, as well as the model's initial prediction of these edges. At early times ($t \approx 15$\,s, Figure~\ref{LSTM}A), the model accurately reproduces the Ti$^{4+}$-dominated spectrum, including the crystal-field splitting of the $t_{2g}$ and $e_g$ contributions. At late times ($t \approx 60$\,s, Figure~\ref{LSTM}B), it tracks the emergence of Ti$^{3+}$ character, namely the merging of white-line features and flattening of the O $K$ edge. There is exceptional agreement between the model's prediction (orange) and the ground-truth (blue) spectra, indicating that it can be used to anticipate changes in electronic state under specific conditions. In practice, this provides an autonomous instrument with the ability to detect slowly evolving phase changes, change course in case of unwanted beam effects, and even extrapolate beyond a present experiment, continually updating its reasoning based on potential trajectories and aiding in discovery of unexpected phase transition pathways.

Modern STEM instruments are inherently multi-dimensional, simultaneously capturing images, spectra, diffraction patterns, and in situ signals, yet most ML models consume only a single modality at a time. Unimodal approaches are limited in their ability to describe physical mechanisms that may manifest weakly, or only partially, in any one signal. STEM-HAADF captures structural contrast but can be insensitive to subtle compositional changes, while EDS resolves composition but may miss fine structural disorder. When both types of change occur simultaneously, as in ion-irradiation of interfaces, neither signal alone is sufficient. To address this, we have developed multi-modal computer vision approaches that combine HAADF and EDS modalities, resolving correlations through community detection, agglomerative clustering, and few-shot classification.\autocite{TerPetrosyan2025npjCM} This approach is a natural extension to unimodal models, but it requires us to codesign model architectures and data selection in tandem to ensure optimal results.

\begin{figure}[H]
\centering \includegraphics[width=\textwidth]{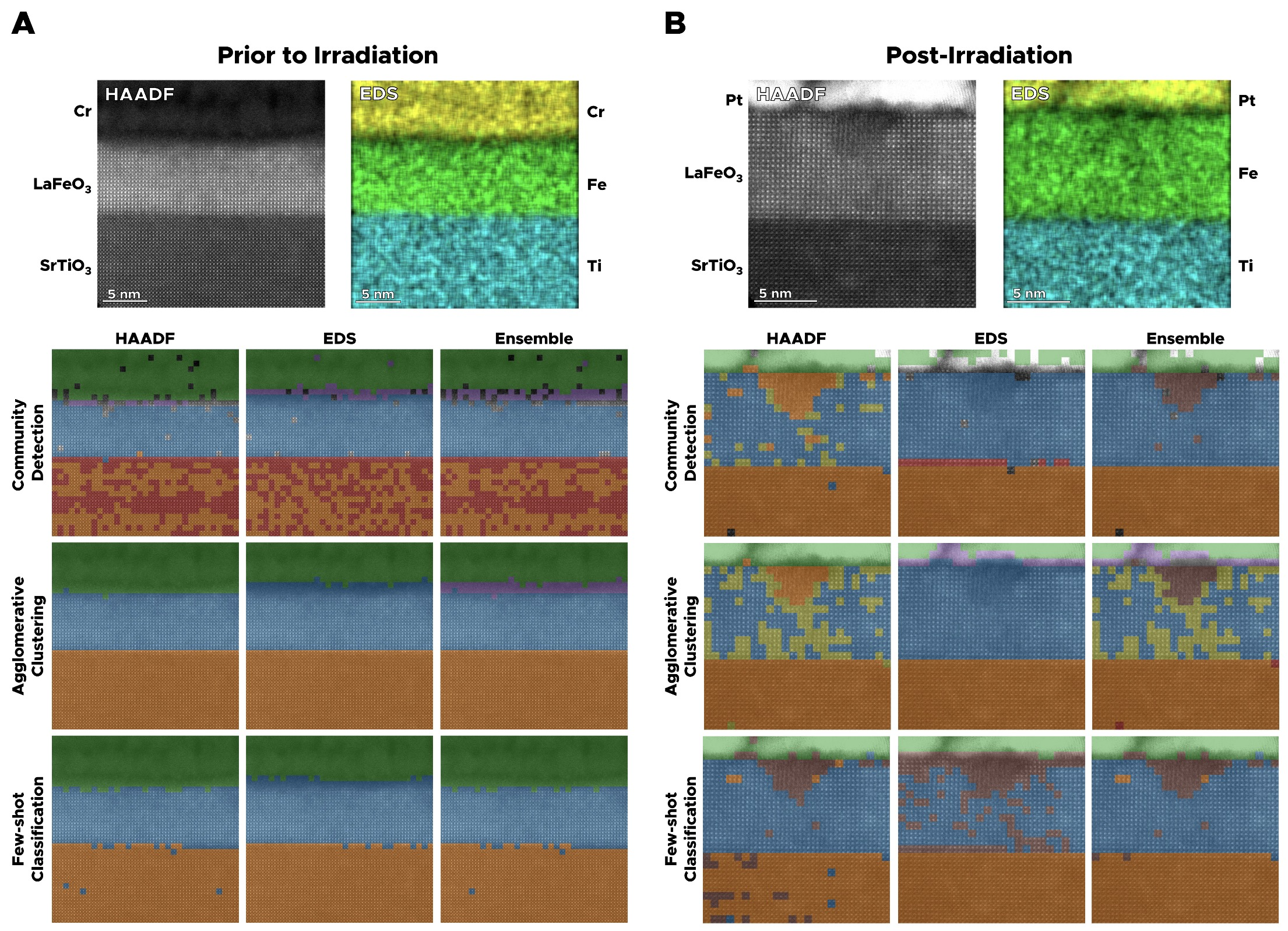}
\caption{\textbf{Multi-modal computer vision reveals irradiation-induced disorder at complex oxide interfaces.} A hybrid machine learning pipeline combining STEM-HAADF and EDS detects structural and chemical signatures of ion-irradiation-induced disorder in epitaxial LaFeO$_3$ (LFO)/SrTiO$_3$ (STO) heterostructures. (\textbf{A})~Prior to irradiation, community detection, agglomerative clustering, and few-shot classification consistently segment the STO, LFO, and capping layers across all modalities. (\textbf{B})~Following irradiation to 0.1\,dpa, single-modality approaches fail to resolve the amorphous front in LFO: HAADF-only models conflate it with the STO substrate, while EDS-only models miss the accompanying compositional changes. The multi-modal ensemble resolves the disordered region across all three classifiers. Reused with permission from reference~\cite{TerPetrosyan2025npjCM} under CC-BY-ND 4.0 license. Copyright 2025 the Authors. \label{multimodal}}
\end{figure}

As shown in Figure~\ref{multimodal}, this approach is particularly powerful for radiation-induced disordering of LaFeO$_3$ (LFO)/SrTiO$_3$ (STO) interfaces. Pre-irradiation (Figure~\ref{multimodal}A), all classification methods correctly separate the STO, LFO, and FIB protective capping layers. After irradiation to 0.1 displacements per atom (dpa, Figure~\ref{multimodal}B), the microstructure has become complex and partially amorphous: HAADF-only models conflate the irradiation-induced amorphous front in LFO with the STO substrate, while EDS-only models miss the subtle compositional changes accompanying amorphization. The multi-modal ensemble resolves what neither modality can alone, revealing local oxygen enrichment and preferential loss of Fe and La cations that inform pathways for radiation-induced disorder. By harnessing multiple modalities, these models reason more effectively and support the development of more accurate physical mechanisms for order-disorder phase transitions.

AI/ML can now practically extract statistical, actionable insight from oxide interfaces. Few-shot learning approaches permit reasoning with limited data in discovery scenarios, yielding quantitative metrics for phase and defect distributions in post hoc analyses. Forecasting models, on the other hand, can act at the time of experiments to guide real-time decision-making, allowing us to intelligently explore the reaction space of a phase transition. Finally, multimodal models harness the full range of signals at our disposal to build much more descriptive and accurate physical models of interfaces. Collectively, these approaches are moving us closer to the day when autonomous, large-scale screening of oxides will enable higher-throughput, more prescriptive discovery and design of materials for diverse applications.\autocite{Olszta2022MicMic,Kalinin2023npjCM,Guinan2025APLML,Guinan2026NatComm}

\section{Concluding Remarks}

In this chapter we have illustrated the range of STEM techniques and
their application to a variety of oxide interfaces and
superlattices. These techniques offer unparalleled spatial resolution
and a wealth of structural, chemical, and compositional information
that can greatly inform models for the behavior of oxides. Advances in
PFIB specimen preparation have expanded the scope
of these measurements, enabling higher-quality, artifact-minimized
lamellae for systems where conventional Ga-FIB preparation would
introduce artifacts. While many
of these methods are mature and well understood, there are still
areas for improvement, particularly in the realm of image
quantification and interpretation---a gap that AI-guided approaches,
including few-shot segmentation, time-series forecasting, and
multi-modal ensemble classification, are now beginning to close. As we look to the future, the integrated use of AI and state-of-the-art STEM imaging will unlock a new level of predictive understanding of oxide interfaces for next-generation technologies.

\clearpage
\section*{Acknowledgments}

The authors acknowledge the foundational contributions of Dr.\ Scott
A.\ Chambers to the original version of this chapter. This work was authored at the National Laboratory of the Rockies, for the U.S. Department of Energy (DOE) under Contract No. DE-AC36-08GO28308. Primary funding for S.R.S. and R.G.\ was provided by the DOE Basic Energy Sciences, Materials Chemistry program to conduct the revision. The original
experimental work on LCO/STO, Fe$_2$O$_3$/Cr$_2$O$_3$, and related
STEM-EDS quantification studies were supported by the
U.S.\ Department of Energy (DOE), Office of Science, Basic Energy
Sciences (BES), Division of Materials Sciences and Engineering under
Award No.\ 10122, and were performed in part at Pacific Northwest
National Laboratory (PNNL), operated for DOE by Battelle. The views
expressed in this article do not necessarily represent the views of
the DOE or the U.S.\ Government. The U.S.\ Government retains and
the publisher, by accepting the article for publication, acknowledges
that the U.S.\ Government retains a nonexclusive, paid-up,
irrevocable, worldwide license to publish or reproduce the published
form of this work, or allow others to do so, for U.S.\ Government
purposes.

\clearpage
\section*{Further Reading}

\subsection*{Books}
\begin{itemize}
  \item Pennycook, S.J.; Nellist, P.D. (Eds.) \textit{Scanning
    Transmission Electron Microscopy}. Springer, 2011.
  \item Egerton, R.F. \textit{Electron Energy-Loss Spectroscopy in
    the Electron Microscope}. Springer, 2011.
  \item Kirkland, E.J. \textit{Advanced Computing in Electron
    Microscopy}. Springer, 2010.
  \item Williams, D.B.; Carter, C.B. \textit{Transmission Electron
    Microscopy}. Springer, 2009.
\end{itemize}

\subsection*{Journal Articles}
\begin{itemize}
  \item Chen, L. et al. {LaAlO$_3$/SrTiO$_3$} heterointerface: 20 years
    and beyond. \textit{Adv.\ Electron.\ Mater.} \textbf{2024},
    \textit{10}, 2300730.
  \item Ophus, C. Quantitative scanning transmission electron microscopy for
    materials science: imaging, diffraction, spectroscopy, and tomography.
    \textit{Annu.\ Rev.\ Mater.\ Res.} \textbf{2023}, \textit{53}, 105--141.
  \item Botifoll, M.; Pinto-Huguet, I.; Arbiol, J. Machine learning
    in electron microscopy for advanced nanocharacterization: current
    developments, available tools and future outlook. \textit{Nanoscale
    Horiz.} \textbf{2022}, \textit{7}, 1427--1477.
  \item Allen, L.J. Simulation in elemental mapping using
    aberration-corrected electron microscopy. \textit{Ultramicroscopy}
    \textbf{2017}, \textit{178}, 110--120.
  \item G\'{a}zquez et al. Applications of STEM-EELS to complex
    oxides. \textit{Mater.\ Sci.\ Semicond.\ Process.}
    \textbf{2017}, \textit{65}, 49--63.
  \item MacLaren, I.; Ramasse, Q.M. Aberration-corrected STEM for
    atomic-resolution studies of functional oxides. \textit{Int.\
    Mater.\ Rev.} \textbf{2014}, \textit{59}, 115--131.
  \item D'Alfonso, A.J. et al. Atomic-resolution chemical mapping
    using EDS. \textit{Phys.\ Rev.\ B} \textbf{2010}, \textit{81},
    100101.
  \item Varela, M. et al. Atomic-resolution imaging of oxidation
    states in manganites. \textit{Phys.\ Rev.\ B} \textbf{2009},
    \textit{79}, 085117.
  \item Muller, D.A. et al. Atomic-scale chemical imaging by
    aberration-corrected microscopy. \textit{Science} \textbf{2008},
    \textit{319}, 1073.
  \item Varela, M. et al. Materials characterization in the
    aberration-corrected STEM. \textit{Annu.\ Rev.\ Mater.\ Res.}
    \textbf{2005}, \textit{35}, 539--569.
\end{itemize}

\subsection*{Selected Works from the Authors}
\begin{itemize}
  \item Guinan, G. et al. Revealing the hidden third dimension of point
    defects in two-dimensional {MXenes}. \textit{Nat.\ Commun.}
    \textbf{2026}, \textit{17}, 3473.
  \item Guinan, G. et al. Mind the gap: Bridging the divide between {AI}
    aspirations and the reality of autonomous microscopy.
    \textit{APL Mach.\ Learn.} \textbf{2025}, \textit{3}, 020903.
  \item Kalinin, S.V. et al. Machine learning for automated experimentation
    in scanning transmission electron microscopy. \textit{npj Comput.\
    Mater.} \textbf{2023}, \textit{9}, 227.
  \item Olszta, M. et al. An automated scanning transmission electron
    microscope guided by sparse data analytics. \textit{Microsc.\
    Microanal.} \textbf{2022}, \textit{28}, 1611--1621.
  \item Spurgeon, S.R. et al. Towards data-driven next-generation
    transmission electron microscopy. \textit{Nat.\ Mater.} \textbf{2021},
    \textit{20}, 274--279.
  \item Spurgeon, S.R. Order-disorder behavior at thin film oxide interfaces.
    \textit{Curr.\ Opin.\ Solid State Mater.\ Sci.} \textbf{2020},
    \textit{24}, 100870.
\end{itemize}

\clearpage
\printbibliography

\end{document}